\documentclass{appolb}
\usepackage{graphicx}
\usepackage{amsmath,amssymb}
\usepackage[dvipsnames]{xcolor}
\usepackage{physics}

\usepackage{mciteplus}

\newcommand{\Yt}{{\mathbf{Y}}}

\newcommand{\xpom}{x_\mathbb{P}}
\newcommand{\xt}{{\mathbf{x}}}

\newcommand{\conj}[1]{\mathop{\overline{#1}}\nolimits}

\newcommand{\Cf}{{C_\mathrm{F}}}

\newcommand{\ud}{\, \mathrm{d}}

\newcommand{\as}{{\alpha_{\mathrm{s}}}}
\newcommand{\rt}{{\mathbf{r}_\perp}}

\newcommand{\bt}{{\mathbf{b}}}

\newcommand{\Deltat}{{\boldsymbol{\Delta}}}

\newcommand{\nc}{{N_\mathrm{c}}}
\newcommand{\jpsi}{$\mathrm{J}/\psi$ }

\newcommand{\cxt}{{\conj{\xt}}}

\newcommand{\contrip}{\conj{0} \conj{1} \conj{2}}

\newcommand{\besk}{{\mathrm{K}}}

\newcommand{\mx}{M_X^2}

\usepackage[breaklinks,colorlinks,citecolor=citcolor,urlcolor=blue,linkcolor=lcolor]{hyperref}
\definecolor{lcolor}{rgb}{0.5,0,0}
\definecolor{citcolor}{rgb}{0,0.3,0.0}

\begin{document}
\title{Diffractive scattering at next-to-leading order in the dipole picture %
\thanks{Presented at the XXIX Cracow Epiphany Conference (January 2023)}%
}
\author{Heikki Mäntysaari
\address{Department of Physics, University of Jyv\"askyl\"a, P.O. Box 35, 40014 University of Jyv\"askyl\"a, Finland, and \\
Helsinki Institute of Physics, P.O. Box 64, 00014 University of Helsinki, Finland}
}

\maketitle
\begin{abstract}
We discuss recent developments towards  next-to-leading order (NLO) accuracy in the dipole picture. We review recent NLO results for exclusive vector meson production and diffractive structure functions, and discuss how it is becoming possible to describe both inclusive and exclusive DIS data simultaneously within the Color Glass Condensate framework at NLO accuracy. These developments will make it possible to probe the gluon saturation phenomena in detail especially when nuclear-DIS data from the EIC become available.
\end{abstract}
  
\section{Introduction}

Diffractive scattering processes are especially powerful probes of the complex QCD dynamics taking place in protons and nuclei at high energies. In these events there can not be a net exchange of color charge. Otherwise one would form a color string that eventually produces hadrons destroying the  gap between the diffractively produced system $X$ (characterized by the invariant mass $M_X^2$, or in exclusive process $X$ can be e.g. a \jpsi meson) and the target. For example the \jpsi production cross section in the collinear factorization based approach is proportional to the squared gluon distribution at leading order~\cite{Brodsky:1994kf} (although the situation becomes more complicated at next-to-leading order~\cite{Eskola:2022vpi}). The second advantage of diffractive processes (especially exclusive vector meson production) is that the total momentum transfer, which is the Fourier conjugate to the impact parameter, is measurable and as such provides access to the spatial distribution of gluons in the target.

The strong sensitivity to the internal structure makes diffractive scattering a powerful probe of non-linear QCD dynamics. This dynamics, which is naturally described in the Color Glass Condensate effective field theory approach to QCD~\cite{Gelis:2010nm}, should manifest itself in the high density region accessed at high energy and/or with heavy nuclei as targets. Probing these non-linear saturation effects is one of the main tasks of the future Electron-Ion Collider at BNL~\cite{AbdulKhalek:2021gbh}, and diffractive processes have an important role in achieving this goal.

\section{Deep inelastic scattering at high energy}

At high energies the Deep Inelastic Scattering (DIS) processes, or similarly photon-mediated interactions in Ultra Peripheral Collisions at RHIC and at the LHC~\cite{Klein:2019qfb} are naturally described in the dipole picture. In the frame where the photon has a large momentum, the lifetime of the partonic fluctuations of the photon state, such as $|q\bar q\rangle$ or $|q\bar q g\rangle$, is much larger than the typical timescale of the interaction. Consequently the scattering amplitude can be factorized to the virtual photon wave function describing  fluctuations to the partonic Fock states, and the dipole-target scattering amplitude describing the eikonal propagation of the quark-antiquark pair in the target color field.

Let us first consider exclusive vector meson production. In the high energy limit the scattering amplitude at leading order (where only the $|q\bar q\rangle$ state contributes)  can be written as
\begin{multline}
\label{eq:lo_vm}
    -i \mathcal{A}^{\gamma^*A \to VA} = \int \dd[2] \bt \dd[2] \rt \frac{\dd z}{4\pi} e^{-i (\bt+(1-2z)/2 \rt) \cdot \mathbf{\Delta}}  \\
    \times \Psi_{\gamma^*}^{q \bar q}(\rt,z) N(\rt,\bt, \xpom) \Psi^{q \bar q*}_V(\rt,z).
\end{multline}
Here the quark and antiquark transverse coordintes are $\bt \pm \rt/2$, $z$ is the fraction of the photon plus momentum carried by the quark and $\Psi_{\gamma^*}$ and $\Psi_V$ refer to the virtual photon and vector meson light front wave functions, and  the vector meson transverse momentum is $\Deltat$.

We emphasize that in the CGC approach the dipole-target scatterling amplitude $N$ is the convenient degree of freedom including all information of the target structure. In particular, the total cross section in DIS can also be obtained from Eq.~\eqref{eq:lo_vm} using the optical theorem by evaluating the amplitude at $\Deltat=0$ and setting $V=\gamma^*$ in which case one obtains the forward elastic scattering amplitude for the $\gamma^* + p \to \gamma^* + p$ scattering. The center-of-mass energy dependence of the dipole amplitude satisfies the Balitsky-Kovchegov equation~\cite{Kovchegov:1999yj,Balitsky:1995ub}. The non-perturbative initial condition for it is typically obtained by fitting the HERA structure function data~\cite{H1:2015ubc}, and excellent descriptions of these measurements have been obtained both at leading~\cite{Lappi:2013zma,Albacete:2010sy} and next-to-leading order accuracy~\cite{Beuf:2020dxl,Hanninen:2022gje}.

At next-to-leading order accuracy one has to include loop corrections to the photon and vector meson wave functions, and consider the $|q\bar q g\rangle$ Fock state that can interact with the target.  The NLO cross section for heavy vector meson production we reported in Refs.~\cite{Mantysaari:2022kdm,Mantysaari:2021ryb}, applying the recently calculated virtual photon NLO wave function~\cite{Beuf:2021srj}. The vector meson wave function is in principle non-perturbative, and in our NLO calculation we described it by including first perturbative corrections to the fully non-relativistic wave function. We additionally included the first relativistic correction at leading order, following Ref.~\cite{Escobedo:2019bxn}. In the limit of high photon virtuality  the meson wave function on the other hand is described in terms of the distribution amplitude, and the corresponding NLO cross section for light meson production at high $Q^2$ we reported in Ref.~\cite{Mantysaari:2022bsp} (the same calculation was previously performed from a slightly different approach in Ref.~\cite{Boussarie:2016bkq}).

When considering diffractive structure functions (diffractive cross sections as a function of e.g. $M_X^2$) there is no need to model the non-perturbative final state wave function which makes such processes theoretically cleaner. In the dipole picture a successful description of the HERA proton diffractive structure function data has been perviously obtained at  leading order~\cite{Kowalski:2008sa}. The successful description of the data however requires one to include a contribution from the produced $q \bar q g$ (color singlet) state which dominates over  the $q\bar q$ production at large $M_X^2$. In the successful phenomenological analysis performed in  Ref.~\cite{Kowalski:2008sa} this contribution was calculated in the high-$Q^2$ limit where the $q\bar q g$ system can be considered as an effective $gg$ dipole. Recently by using the exact light cone wave function describing the $\gamma^* \to q\bar q g$ splitting~\cite{Beuf:2017bpd} (neglecting the quark masses) and performing the final state phase space integral exactly we calculated the leading $q\bar q g$ contribution to diffractive structure functions in exact kinematics~\cite{Beuf:2022kyp}, and showed that in the high-$Q^2$ limit we obtain the previously known so called ``Wusthoff result''~\cite{Wusthoff:1997fz}. This contribution where the $q\bar q g$ system interacts with the shockwave is itself finite at fixed $M_X^2$ and reads
\begin{multline}
	F_{T, \, q \bar q  g}^{\textrm{D}(4) \, \textrm{NLO}} (\xpom, Q^2, M_X^2, t)
	=
	2 \nc Q^2
	%				\frac{\as \cf}{(2\pi)^6} % 1/(2\pi)^6 goes into the d^2 xt phase spaces
	\as \Cf
	\sum_f e_f^2
	\int_0^1 \frac{\ud z_0}{z_0}
	\int_0^1 \frac{\ud z_1}{z_1}
	\frac{1}{z_2}
	%\int_0^1 \frac{\ud z_2}{z_2}
% 	\nonumber
% 	\\
% 	& \qquad \times
	%\delta(z_0 \! + \! z_1 \! + \! z_2 \! - \! 1)
	\\
	\qquad \times
	\int_{\xt_0} \int_{\xt_1} \int_{\xt_2} \int_{\cxt_0} \int_{\cxt_1} \int_{\cxt_2}
	{\cal I}_{\mx}^{(3)} {\cal I}_{\Deltat}^{(3)}
% 	\nonumber
% 	\\
% 	& \qquad \times
	\frac{z_0 z_1 Q^2}{X_{012} X_{\conj{0}\conj{1}\conj{2} }}
	%			\\
	%			\times
	\besk_1\left(QX_{012}\right) \besk_1\left(Q X_{\conj{0}\conj{1}\conj{2} }\right)
	\nonumber
	\\
	\qquad \times
	\Big\lbrace
	\Upsilon^{(|a|^2)}_{\textrm{reg.}} + \Upsilon^{(|b|^2)}_{\textrm{reg.}} 
%				\nonumber
%				\\
%				& \qquad
	+ \Upsilon^{(a')}_{\textrm{inst.}} + \Upsilon^{(b')}_{\textrm{inst.}} + \Upsilon^{(ab)}_{\textrm{interf.}}
	\Big\rbrace
%				\nonumber
%				\\
%				& \qquad \times
%				\left[S_{\contrip}^{(3)\dagger} - 1\right] \left[S_{012}^{(3)} - 1\right]
	 N_{\contrip}  N_{012}
	\label{eq:ddis-FT-qqbarg-nlo}
\end{multline}
for the transverse structure function (results for the longitudinal cross section are also reported in Ref~\cite{Beuf:2022kyp}).
Here the quark, antiquark and gluon transverse coordinates are $\xt_0,\xt_1$ and $\xt_2$ and coordinates with bar refer to the conjugate amplitude, $Q^2X_{012}^2$ is the $q\bar q g$ formation time divided by the lifetime of the virtual photon that forms the $q\bar q g$ system and $z_2=1-z_1-z_0$ is the gluon longitudinal momentum fraction. The interaction with the target is described by $N_{012}$ which is the $q\bar q g$-target scattering amplitude that can be written in terms of dipole-target amplitudes using the Fierz identity. Detailed expressions for the $\Upsilon$ terms that originate from the square of the photon wave function can be found from Ref.~\cite{Beuf:2022kyp}. The transfer function $ {\cal I}_{\mx}^{(3)}$ describing the formation of a system with given invariant mass from the coordinate space $q\bar q g$ system reads
\begin{equation}
\label{eq:IMX3}
    {\cal I}_{\mx}^{(3)} =
        2 \frac{z_0 z_1 z_2}{(4 \pi)^2} \frac{\mx}{Y_{012}} \mathrm{J}_1 \left( \mx \abs{\Yt_{012}} \right),
\end{equation}
where
\begin{equation}
    \Yt_{012}^2 =
        z_0 z_1 \left( \xt_{\overline 0 0} - \xt_{\overline 1 1} \right)^2 +
        z_1 z_2 \left( \xt_{\overline 2 2} - \xt_{\overline 1 1} \right)^2 +
        z_0 z_2 \left( \xt_{\overline 2 2} - \xt_{\overline 0 0} \right)^2.
\end{equation}
Note that $\Yt_{012}^2$ does not depend on the center-of-mass of the $q\bar q g$ system $\bt = z_0 \xt_0 + z_1 \xt_1 + z_2 \xt_2$ (or on $\overline \bt$) which is the Fourier conjugate to the momentumn transfer $\Deltat$, and when integrating over the momentum transfer the other transfer function ${\cal I}_{\Deltat}^{(3)}$ sets $\bt=\conj \bt$.

In order to obtain the full NLO result for diffractive structure functions one needs to add all possible loop corrections and gluon emission after the shockwave. These contributions are separately divergent, but final result for the cross section is of course finite.
These results we will report in a separate publication in the future. 
When this calculation is finished, it becomes possible to simultaneously calculate exclusive vector meson production and diffractive structure functions in the dipole picture at NLO consistently with inclusive structure function data. This is crucial to test the gluon saturation picture of CGC especially in the EIC era with precise nuclear-DIS data.

\section{Numerical results for exclusive vector meson production}

 \begin{figure}[tb]
 \begin{minipage}{.49\textwidth}
 \centering
 		\includegraphics[width=1\textwidth]{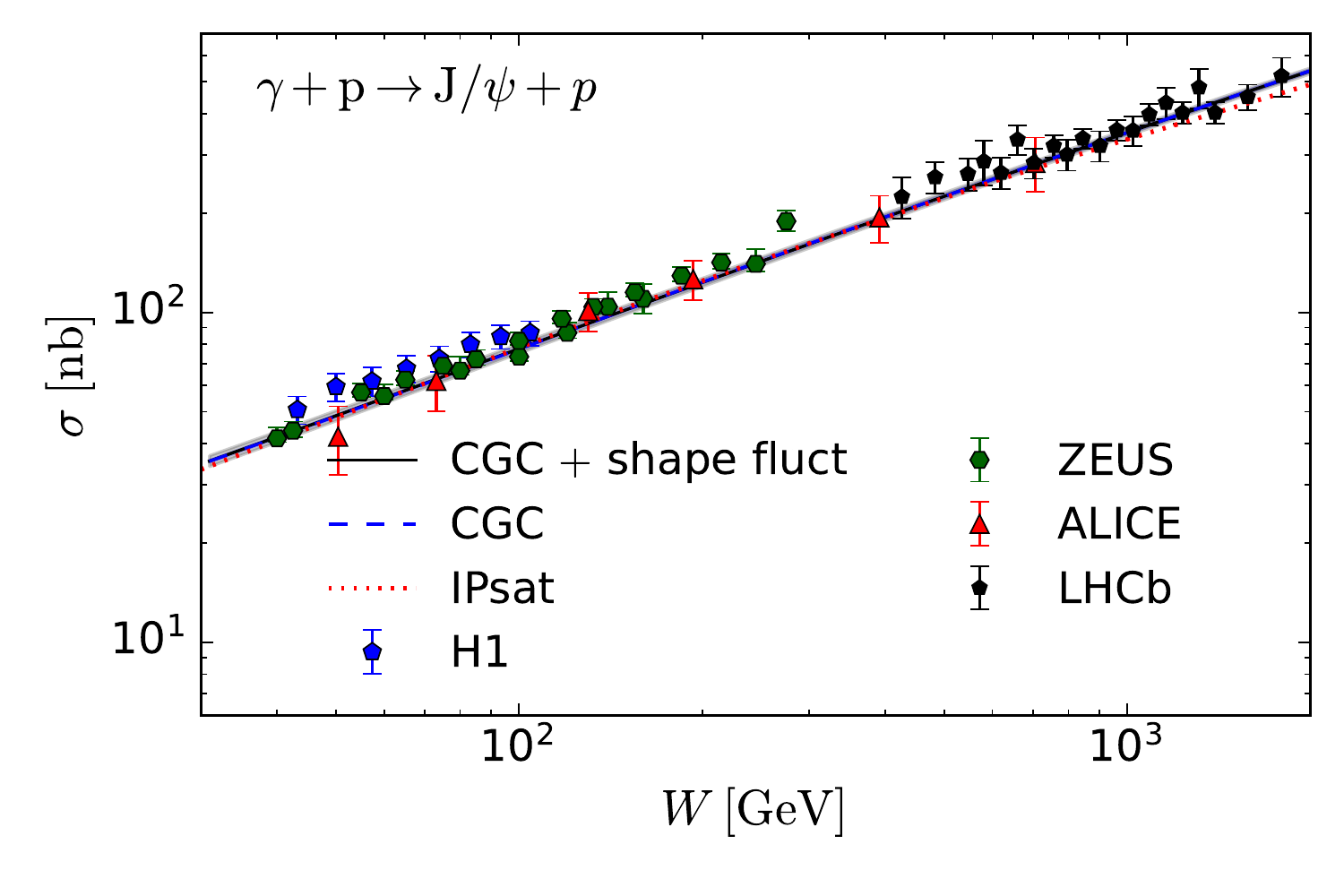}  %width=\columnwidth
 				\caption{Coherent \jpsi production cross section as a function of center-of-mass energy $W$ from Ref.~\cite{Mantysaari:2022sux}.}  
 			\label{fig:gamma_jpsi_coh}

 \end{minipage}
 \quad
 \begin{minipage}{.49\textwidth}
 \centering
 \includegraphics[width=\textwidth]{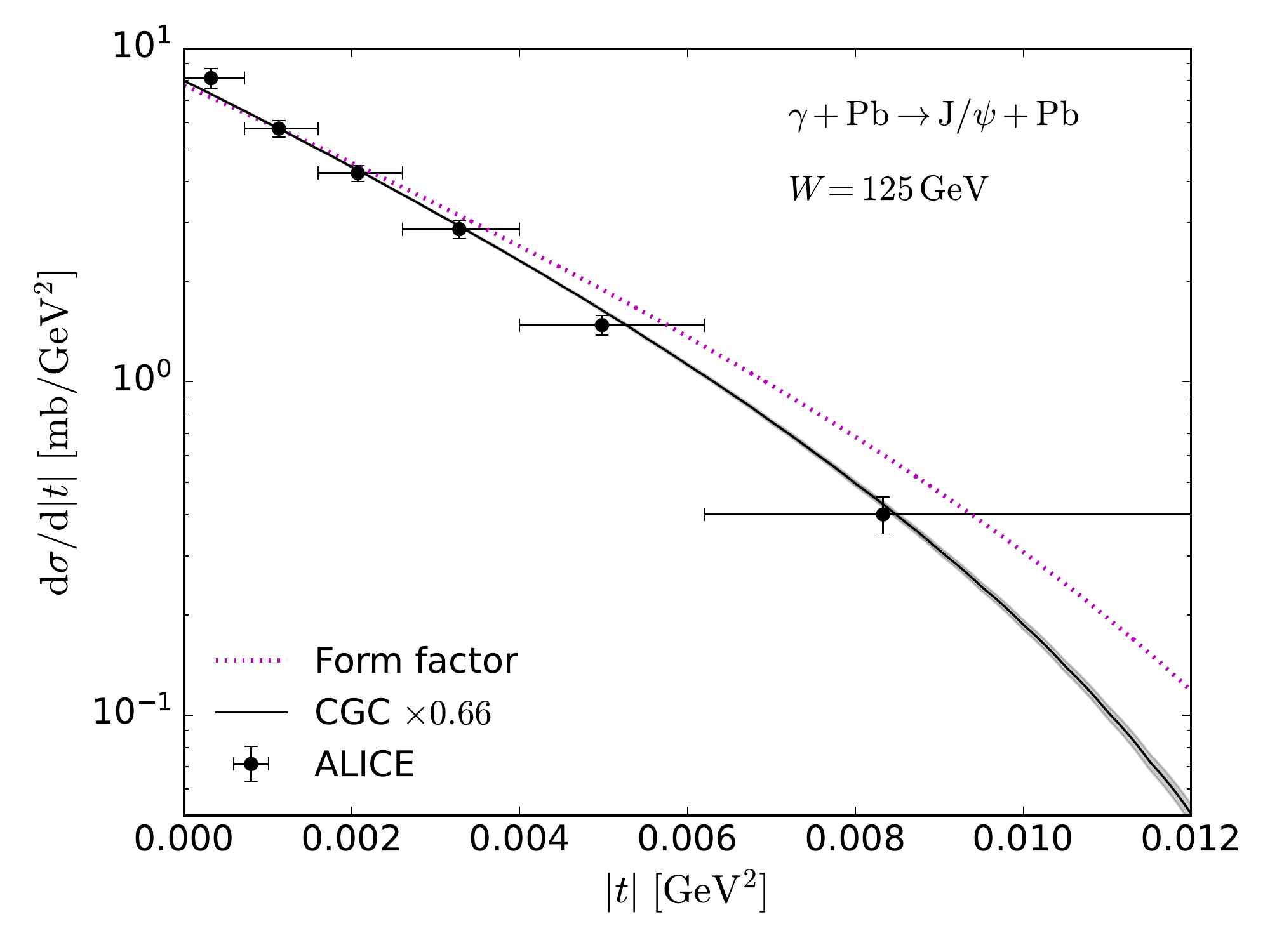} 
 				\caption{Differential coherent \jpsi production cross section in photon-nucleus collisions  compared to the ALICE data~\cite{Acharya:2021bnz}. Figure adapted from Ref.~\cite{Mantysaari:2022sux}.}  
 			\label{fig:alice_t}
 \end{minipage}
 \end{figure}

Let us next highlight some selected numerical results for exclusive vector meson production. To demonstrate the successful phenomenology at leading order we show in Fig.~\ref{fig:gamma_jpsi_coh} the $t$-integrated coherent \jpsi production cross section as a function of center-of-mass energy $W$. The CGC results from Ref.~\cite{Mantysaari:2022sux} are found to describe well  the available high-energy data. In principle saturation effects could be expected to slow down the $W$ dependence at high energies/densities, but no significant effect is predicted based on the  small-$x$ evolution of the proton structure included in the calculation.

As non-linear effects are enhanced in heavy nuclei we also calculate coherent \jpsi production in photon-nucleus collisions. In ultra peripheral collisions at the LHC significant nuclear suppression is observed~\cite{ALICE:2021gpt}. As discussed in~\cite{Mantysaari:2022sux}, we typically overestimate the nuclear cross sections, i.e. the CGC calculations tend to underestimate the amount of suppression especially close to midrapidity in LHC kinematics. In addition to overall normalization, it is also possible to look at more differential observables such as $t$ distributions that probe the spatial distribution of small-$x$ gluons in the target. The calculated spectra and comparison to the ALICE data are shown in Fig.~\ref{fig:alice_t}. When the normalization is fixed by the data, we find that the  CGC calculation including non-linear effects describes the shape of the ALICE data well. On the other hand if non-linear effects are neglected (corresponding to the curve \emph{Form factor} in Fig.~\ref{fig:alice_t}) a clearly too hard spectrum is obtained. This result can be interpreted such that strong non-linear effects close to the center of the nucleus modify the spatial density profile to be different from the Woods-Saxon distribution which is an input to the calculation and from which the positions of individual nucleons are sampled.

% %uncomment the following lines to place a figure
% \begin{figure}[tb]
% \centerline{%
% \includegraphics[width=10.8cm]{posterior.pdf}}
% \caption{Posterior distribution of the model parameters. }
% \label{Fig:posterior}
% \end{figure}

 \begin{figure}[tb]
 \begin{minipage}{.49\textwidth}
 \centering
 		\includegraphics[width=\textwidth]{xs-rel-W_90}  %width=\columnwidth
 				\caption{Coherent \jpsi production at fixed $W=75\,\mathrm{GeV}$ from Ref.~\cite{Mantysaari:2022kdm}.}  
 			\label{fig:nlo_jpsi}

 \end{minipage}
 \quad
 \begin{minipage}{.49\textwidth}
 \centering
 \includegraphics[width=\textwidth]{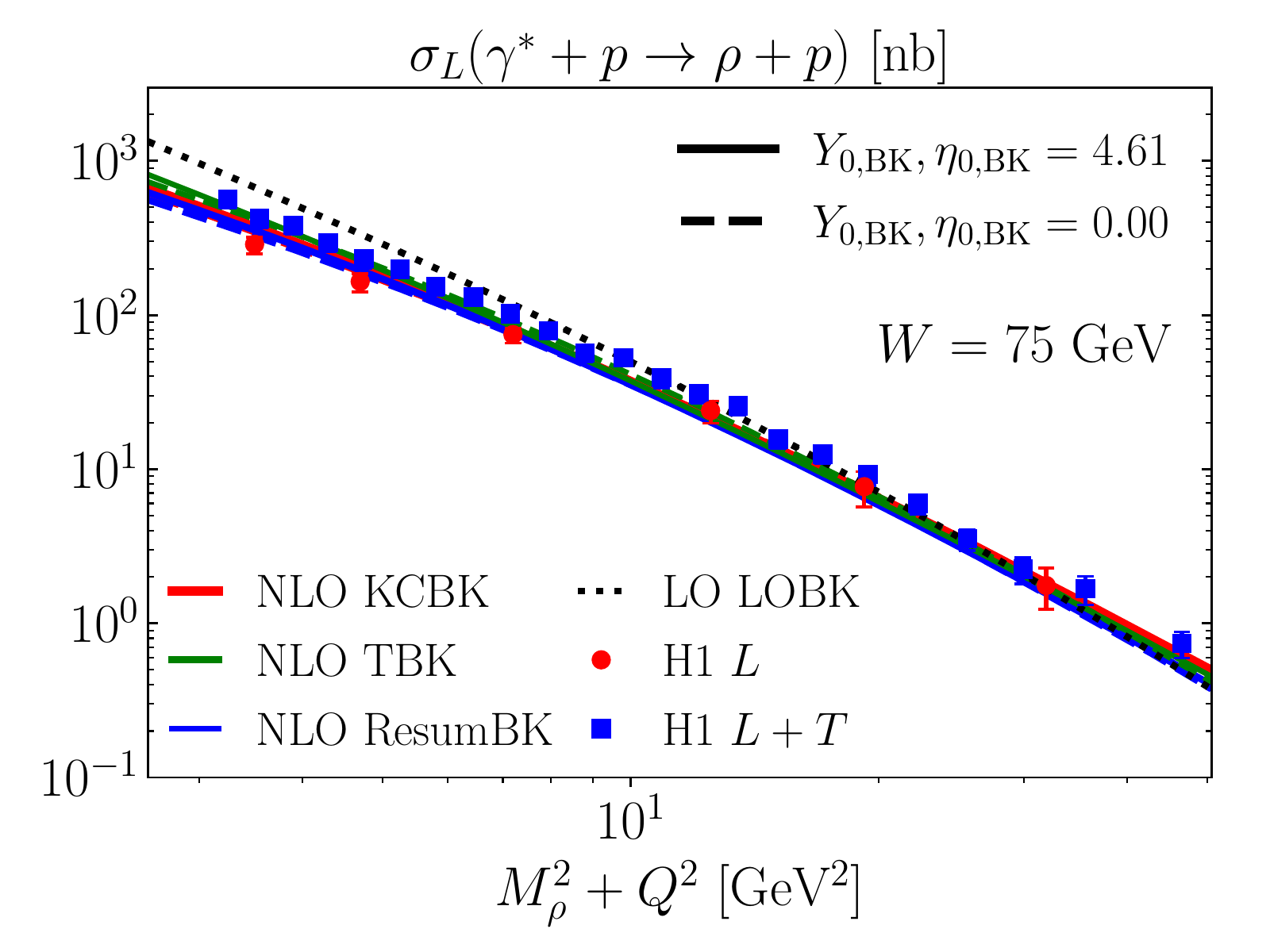} 
 				\caption{Coherent $\rho$ production at fixed $W=75\,\mathrm{GeV}$ from Ref.~\cite{Mantysaari:2022bsp}.}  
 			\label{fig:nlo_rho}
 \end{minipage}
 \end{figure}

 When vector meson production is computed at next-to-leading order it is also crucial to use a corresponding NLO-evolved dipole-proton scattering amplitude, with the initial condition for the BK evolution fitted to the data at NLO accuracy. In Refs.~\cite{Mantysaari:2022kdm,Mantysaari:2022bsp} we used different fits reported in Ref.~\cite{Beuf:2020dxl} and the deviations between the results shown in Figs.~\ref{fig:nlo_jpsi} and~\ref{fig:nlo_rho} correspond to the remaining model uncertainty. We show results separately for \jpsi  (Fig.~\ref{fig:nlo_jpsi} based on  Ref.~\cite{Mantysaari:2022kdm}) and $\rho$ (Fig.~\ref{fig:nlo_rho} based on Ref.~\cite{Mantysaari:2022bsp}) production. When calculating the \jpsi production we also include the first relativistic correction to the \jpsi wave function (in our power counting $v^2\sim \as$ where $v$ is the heavy quark velocity in the \jpsi rest frame). This is required  to obtain a good description of the data also at low $Q^2$. With the first relativistic correction included we find a good description of the HERA vector meson production data, and also note that in this kinematics the NLO corrections are moderate.
 
Similarly a good desription of the $\rho$ production cross section in the high-virtuality region is demonstrated in Fig.~\ref{fig:nlo_rho}. We only calculate the contribution where a longitudinal $\rho$ is formed by a longitudinal photon, which dominates at high $Q^2$. Our results also agree with the experimental data at low $Q^2$, but we emphasize that the calculation is not valid there and also a numerically important transverse component is missing. In the high-virtuality region where the calculation is valid an excellent description of the H1 data is obtained.

\section{Conclusions}

The Color Glass Condensate framework has been extensively used to successfully  describe many different scattering processes sensitive to the proton and nuclear small-$x$ structure. However, in order to accurately probe the non-linear QCD dynamics in protons and especially in heavy nuclei it is necessary to promote the field to the NLO accuracy. There has been a lot of progress in that direction recently. In this Talk, we summarized recent developments towards the NLO accuracy related to inclusive and diffractive structure functions and exclusive vector meson production. These developments have made it possible to describe, for the first time in dipole picture, simultaneously the total cross section and heavy quark production in DIS in HERA kinematics. Cross sections for exclusive heavy and light vector meson production have been calculated at NLO, and a subset of NLO corrections to the diffractive structure functions is also available in the literature. These developments should make global analyses of precise small-$x$ DIS data possible in the future, which will make it possible to study in the detail the gluon saturation phenomena in current colliders and especially at the EIC.

\subsection*{Acknowledgments}
H.M. is supported by the Academy of Finland, the Centre of Excellence in Quark Matter, and projects 338263 and 346567, and under the European Union’s Horizon 2020 research and innovation programme by the European Research Council (ERC, grant agreement No. ERC-2018-ADG-835105 YoctoLHC) and by the STRONG-2020 project (grant agreement No. 824093).

\bibliographystyle{JHEP-2modM}
\bibliography{refs}

\ifx\mcitethebibliography\mciteundefinedmacro
\PackageError{JHEP-2modM.bst}{mciteplus.sty has not been loaded}
{This bibstyle requires the use of the mciteplus package.}\fi
\providecommand{\href}[2]{#2}
\begingroup\raggedright\begin{mcitethebibliography}{10}

\bibitem{Brodsky:1994kf}
S.~J. Brodsky, L.~Frankfurt, J.~F. Gunion, A.~H. Mueller and M.~Strikman,
  \mbox{}  \href{http://dx.doi.org/10.1103/PhysRevD.50.3134}{{\em Phys. Rev. D}
  {\bf 50} (1994) 3134} [\href{http://arXiv.org/abs/hep-ph/9402283}{{\tt
  arXiv:hep-ph/9402283}}]\relax
\mciteBstWouldAddEndPuncttrue
\mciteSetBstMidEndSepPunct{\mcitedefaultmidpunct}
{\mcitedefaultendpunct}{\mcitedefaultseppunct}\relax
\EndOfBibitem
\bibitem{Eskola:2022vpi}
K.~J. Eskola, C.~A. Flett, V.~Guzey, T.~L\"oyt\"ainen and H.~Paukkunen,
  \mbox{}  \href{http://dx.doi.org/10.1103/PhysRevC.106.035202}{{\em Phys. Rev.
  C} {\bf 106} (2022)no.~3 035202} [\href{http://arXiv.org/abs/2203.11613}{{\tt
  arXiv:2203.11613 [hep-ph]}}]\relax
\mciteBstWouldAddEndPuncttrue
\mciteSetBstMidEndSepPunct{\mcitedefaultmidpunct}
{\mcitedefaultendpunct}{\mcitedefaultseppunct}\relax
\EndOfBibitem
\bibitem{Gelis:2010nm}
F.~Gelis, E.~Iancu, J.~Jalilian-Marian and R.~Venugopalan,  \mbox{}
  \href{http://dx.doi.org/10.1146/annurev.nucl.010909.083629}{{\em Ann. Rev.
  Nucl. Part. Sci.} {\bf 60} (2010) 463}
  [\href{http://arXiv.org/abs/1002.0333}{{\tt arXiv:1002.0333 [hep-ph]}}]\relax
\mciteBstWouldAddEndPuncttrue
\mciteSetBstMidEndSepPunct{\mcitedefaultmidpunct}
{\mcitedefaultendpunct}{\mcitedefaultseppunct}\relax
\EndOfBibitem
\bibitem{AbdulKhalek:2021gbh}
R.~Abdul~Khalek {\em et.~al.},  \mbox{}
  \href{http://dx.doi.org/10.1016/j.nuclphysa.2022.122447}{{\em Nucl. Phys. A}
  {\bf 1026} (2022) 122447} [\href{http://arXiv.org/abs/2103.05419}{{\tt
  arXiv:2103.05419 [physics.ins-det]}}]\relax
\mciteBstWouldAddEndPuncttrue
\mciteSetBstMidEndSepPunct{\mcitedefaultmidpunct}
{\mcitedefaultendpunct}{\mcitedefaultseppunct}\relax
\EndOfBibitem
\bibitem{Klein:2019qfb}
S.~R. Klein and H.~M\"antysaari,  \mbox{}
  \href{http://dx.doi.org/10.1038/s42254-019-0107-6}{{\em Nature Rev. Phys.}
  {\bf 1} (2019)no.~11 662} [\href{http://arXiv.org/abs/1910.10858}{{\tt
  arXiv:1910.10858 [hep-ex]}}]\relax
\mciteBstWouldAddEndPuncttrue
\mciteSetBstMidEndSepPunct{\mcitedefaultmidpunct}
{\mcitedefaultendpunct}{\mcitedefaultseppunct}\relax
\EndOfBibitem
\bibitem{Kovchegov:1999yj}
Y.~V. Kovchegov,  \mbox{}
  \href{http://dx.doi.org/10.1103/PhysRevD.60.034008}{{\em Phys. Rev. D} {\bf
  60} (1999) 034008} [\href{http://arXiv.org/abs/hep-ph/9901281}{{\tt
  arXiv:hep-ph/9901281}}]\relax
\mciteBstWouldAddEndPuncttrue
\mciteSetBstMidEndSepPunct{\mcitedefaultmidpunct}
{\mcitedefaultendpunct}{\mcitedefaultseppunct}\relax
\EndOfBibitem
\bibitem{Balitsky:1995ub}
I.~Balitsky,  \mbox{}
  \href{http://dx.doi.org/10.1016/0550-3213(95)00638-9}{{\em Nucl. Phys. B}
  {\bf 463} (1996) 99} [\href{http://arXiv.org/abs/hep-ph/9509348}{{\tt
  arXiv:hep-ph/9509348}}]\relax
\mciteBstWouldAddEndPuncttrue
\mciteSetBstMidEndSepPunct{\mcitedefaultmidpunct}
{\mcitedefaultendpunct}{\mcitedefaultseppunct}\relax
\EndOfBibitem
\bibitem{H1:2015ubc}
{\bf H1, ZEUS} collaboration, H.~Abramowicz {\em et.~al.},  \mbox{}
  \href{http://dx.doi.org/10.1140/epjc/s10052-015-3710-4}{{\em Eur. Phys. J. C}
  {\bf 75} (2015)no.~12 580} [\href{http://arXiv.org/abs/1506.06042}{{\tt
  arXiv:1506.06042 [hep-ex]}}]\relax
\mciteBstWouldAddEndPuncttrue
\mciteSetBstMidEndSepPunct{\mcitedefaultmidpunct}
{\mcitedefaultendpunct}{\mcitedefaultseppunct}\relax
\EndOfBibitem
\bibitem{Lappi:2013zma}
T.~Lappi and H.~M\"antysaari,  \mbox{}
  \href{http://dx.doi.org/10.1103/PhysRevD.88.114020}{{\em Phys. Rev. D} {\bf
  88} (2013) 114020} [\href{http://arXiv.org/abs/1309.6963}{{\tt
  arXiv:1309.6963 [hep-ph]}}]\relax
\mciteBstWouldAddEndPuncttrue
\mciteSetBstMidEndSepPunct{\mcitedefaultmidpunct}
{\mcitedefaultendpunct}{\mcitedefaultseppunct}\relax
\EndOfBibitem
\bibitem{Albacete:2010sy}
J.~L. Albacete, N.~Armesto, J.~G. Milhano, P.~Quiroga-Arias and C.~A. Salgado,
  \mbox{}  \href{http://dx.doi.org/10.1140/epjc/s10052-011-1705-3}{{\em Eur.
  Phys. J. C} {\bf 71} (2011) 1705} [\href{http://arXiv.org/abs/1012.4408}{{\tt
  arXiv:1012.4408 [hep-ph]}}]\relax
\mciteBstWouldAddEndPuncttrue
\mciteSetBstMidEndSepPunct{\mcitedefaultmidpunct}
{\mcitedefaultendpunct}{\mcitedefaultseppunct}\relax
\EndOfBibitem
\bibitem{Beuf:2020dxl}
G.~Beuf, H.~H\"anninen, T.~Lappi and H.~M\"antysaari,  \mbox{}
  \href{http://dx.doi.org/10.1103/PhysRevD.102.074028}{{\em Phys. Rev. D} {\bf
  102} (2020) 074028} [\href{http://arXiv.org/abs/2007.01645}{{\tt
  arXiv:2007.01645 [hep-ph]}}]\relax
\mciteBstWouldAddEndPuncttrue
\mciteSetBstMidEndSepPunct{\mcitedefaultmidpunct}
{\mcitedefaultendpunct}{\mcitedefaultseppunct}\relax
\EndOfBibitem
\bibitem{Hanninen:2022gje}
H.~H\"anninen, H.~M\"antysaari, R.~Paatelainen and J.~Penttala,  \mbox{}
  \href{http://arXiv.org/abs/2211.03504}{{\tt arXiv:2211.03504 [hep-ph]}}\relax
\mciteBstWouldAddEndPuncttrue
\mciteSetBstMidEndSepPunct{\mcitedefaultmidpunct}
{\mcitedefaultendpunct}{\mcitedefaultseppunct}\relax
\EndOfBibitem
\bibitem{Mantysaari:2022kdm}
H.~M\"antysaari and J.~Penttala,  \mbox{}
  \href{http://dx.doi.org/10.1007/JHEP08(2022)247}{{\em JHEP} {\bf 08} (2022)
  247} [\href{http://arXiv.org/abs/2204.14031}{{\tt arXiv:2204.14031
  [hep-ph]}}]\relax
\mciteBstWouldAddEndPuncttrue
\mciteSetBstMidEndSepPunct{\mcitedefaultmidpunct}
{\mcitedefaultendpunct}{\mcitedefaultseppunct}\relax
\EndOfBibitem
\bibitem{Mantysaari:2021ryb}
H.~M\"antysaari and J.~Penttala,  \mbox{}
  \href{http://dx.doi.org/10.1016/j.physletb.2021.136723}{{\em Phys. Lett. B}
  {\bf 823} (2021) 136723} [\href{http://arXiv.org/abs/2104.02349}{{\tt
  arXiv:2104.02349 [hep-ph]}}]\relax
\mciteBstWouldAddEndPuncttrue
\mciteSetBstMidEndSepPunct{\mcitedefaultmidpunct}
{\mcitedefaultendpunct}{\mcitedefaultseppunct}\relax
\EndOfBibitem
\bibitem{Beuf:2021srj}
G.~Beuf, T.~Lappi and R.~Paatelainen,  \mbox{}
  \href{http://dx.doi.org/10.1103/PhysRevLett.129.072001}{{\em Phys. Rev.
  Lett.} {\bf 129} (2022)no.~7 072001}
  [\href{http://arXiv.org/abs/2112.03158}{{\tt arXiv:2112.03158
  [hep-ph]}}]\relax
\mciteBstWouldAddEndPuncttrue
\mciteSetBstMidEndSepPunct{\mcitedefaultmidpunct}
{\mcitedefaultendpunct}{\mcitedefaultseppunct}\relax
\EndOfBibitem
\bibitem{Escobedo:2019bxn}
M.~A. Escobedo and T.~Lappi,  \mbox{}
  \href{http://dx.doi.org/10.1103/PhysRevD.101.034030}{{\em Phys. Rev. D} {\bf
  101} (2020)no.~3 034030} [\href{http://arXiv.org/abs/1911.01136}{{\tt
  arXiv:1911.01136 [hep-ph]}}]\relax
\mciteBstWouldAddEndPuncttrue
\mciteSetBstMidEndSepPunct{\mcitedefaultmidpunct}
{\mcitedefaultendpunct}{\mcitedefaultseppunct}\relax
\EndOfBibitem
\bibitem{Mantysaari:2022bsp}
H.~M\"antysaari and J.~Penttala,  \mbox{}
  \href{http://dx.doi.org/10.1103/PhysRevD.105.114038}{{\em Phys. Rev. D} {\bf
  105} (2022)no.~11 114038} [\href{http://arXiv.org/abs/2203.16911}{{\tt
  arXiv:2203.16911 [hep-ph]}}]\relax
\mciteBstWouldAddEndPuncttrue
\mciteSetBstMidEndSepPunct{\mcitedefaultmidpunct}
{\mcitedefaultendpunct}{\mcitedefaultseppunct}\relax
\EndOfBibitem
\bibitem{Boussarie:2016bkq}
R.~Boussarie, A.~V. Grabovsky, D.~Y. Ivanov, L.~Szymanowski and S.~Wallon,
  \mbox{}  \href{http://dx.doi.org/10.1103/PhysRevLett.119.072002}{{\em Phys.
  Rev. Lett.} {\bf 119} (2017)no.~7 072002}
  [\href{http://arXiv.org/abs/1612.08026}{{\tt arXiv:1612.08026
  [hep-ph]}}]\relax
\mciteBstWouldAddEndPuncttrue
\mciteSetBstMidEndSepPunct{\mcitedefaultmidpunct}
{\mcitedefaultendpunct}{\mcitedefaultseppunct}\relax
\EndOfBibitem
\bibitem{Kowalski:2008sa}
H.~Kowalski, T.~Lappi, C.~Marquet and R.~Venugopalan,  \mbox{}
  \href{http://dx.doi.org/10.1103/PhysRevC.78.045201}{{\em Phys. Rev. C} {\bf
  78} (2008) 045201} [\href{http://arXiv.org/abs/0805.4071}{{\tt
  arXiv:0805.4071 [hep-ph]}}]\relax
\mciteBstWouldAddEndPuncttrue
\mciteSetBstMidEndSepPunct{\mcitedefaultmidpunct}
{\mcitedefaultendpunct}{\mcitedefaultseppunct}\relax
\EndOfBibitem
\bibitem{Beuf:2017bpd}
G.~Beuf,  \mbox{}  \href{http://dx.doi.org/10.1103/PhysRevD.96.074033}{{\em
  Phys. Rev. D} {\bf 96} (2017)no.~7 074033}
  [\href{http://arXiv.org/abs/1708.06557}{{\tt arXiv:1708.06557
  [hep-ph]}}]\relax
\mciteBstWouldAddEndPuncttrue
\mciteSetBstMidEndSepPunct{\mcitedefaultmidpunct}
{\mcitedefaultendpunct}{\mcitedefaultseppunct}\relax
\EndOfBibitem
\bibitem{Beuf:2022kyp}
G.~Beuf, H.~H\"anninen, T.~Lappi, Y.~Mulian and H.~M\"antysaari,  \mbox{}
  \href{http://dx.doi.org/10.1103/PhysRevD.106.094014}{{\em Phys. Rev. D} {\bf
  106} (2022)no.~9 094014} [\href{http://arXiv.org/abs/2206.13161}{{\tt
  arXiv:2206.13161 [hep-ph]}}]\relax
\mciteBstWouldAddEndPuncttrue
\mciteSetBstMidEndSepPunct{\mcitedefaultmidpunct}
{\mcitedefaultendpunct}{\mcitedefaultseppunct}\relax
\EndOfBibitem
\bibitem{Wusthoff:1997fz}
M.~Wusthoff,  \mbox{}  \href{http://dx.doi.org/10.1103/PhysRevD.56.4311}{{\em
  Phys. Rev. D} {\bf 56} (1997) 4311}
  [\href{http://arXiv.org/abs/hep-ph/9702201}{{\tt
  arXiv:hep-ph/9702201}}]\relax
\mciteBstWouldAddEndPuncttrue
\mciteSetBstMidEndSepPunct{\mcitedefaultmidpunct}
{\mcitedefaultendpunct}{\mcitedefaultseppunct}\relax
\EndOfBibitem
\bibitem{Mantysaari:2022sux}
H.~M\"antysaari, F.~Salazar and B.~Schenke,  \mbox{}
  \href{http://dx.doi.org/10.1103/PhysRevD.106.074019}{{\em Phys. Rev. D} {\bf
  106} (2022)no.~7 074019} [\href{http://arXiv.org/abs/2207.03712}{{\tt
  arXiv:2207.03712 [hep-ph]}}]\relax
\mciteBstWouldAddEndPuncttrue
\mciteSetBstMidEndSepPunct{\mcitedefaultmidpunct}
{\mcitedefaultendpunct}{\mcitedefaultseppunct}\relax
\EndOfBibitem
\bibitem{Acharya:2021bnz}
{\bf ALICE} collaboration, S.~Acharya {\em et.~al.},  \mbox{}
  \href{http://dx.doi.org/10.1016/j.physletb.2021.136280}{{\em Phys. Lett. B}
  {\bf 817} (2021) 136280} [\href{http://arXiv.org/abs/2101.04623}{{\tt
  arXiv:2101.04623 [nucl-ex]}}]\relax
\mciteBstWouldAddEndPuncttrue
\mciteSetBstMidEndSepPunct{\mcitedefaultmidpunct}
{\mcitedefaultendpunct}{\mcitedefaultseppunct}\relax
\EndOfBibitem
\bibitem{ALICE:2021gpt}
{\bf ALICE} collaboration, S.~Acharya {\em et.~al.},  \mbox{}
  \href{http://dx.doi.org/10.1140/epjc/s10052-021-09437-6}{{\em Eur. Phys. J.
  C} {\bf 81} (2021)no.~8 712} [\href{http://arXiv.org/abs/2101.04577}{{\tt
  arXiv:2101.04577 [nucl-ex]}}]\relax
\mciteBstWouldAddEndPuncttrue
\mciteSetBstMidEndSepPunct{\mcitedefaultmidpunct}
{\mcitedefaultendpunct}{\mcitedefaultseppunct}\relax
\EndOfBibitem
\end{mcitethebibliography}\endgroup

\end{document}